\documentclass{cimento}

\usepackage{graphicx} 
\title{The STAR $W$ Program at RHIC}
\author{J.~R.~Stevens\from{ins:x}, for the STAR Collaboration}
\instlist{\inst{ins:x} Massachusetts Institute of Technology, Cambridge, MA, USA}
\begin{document}

\maketitle

\begin{abstract}

The production of $W$ bosons in polarized $p+p$ collisions at RHIC provides an excellent tool to probe the proton's sea quark distributions.   At leading order $W^{-(+)}$ bosons are produced in $\bar{u}+d\,(\bar{d}+u)$ collisions, and parity-violating single-spin asymmetries measured in longitudinally polarized $p+p$ collisions give access to the flavor-separated light quark and antiquark helicity distributions.  In this proceedings we report preliminary results for the single-spin asymmetry, $A_L$ from data collected in 2012 by the STAR experiment at RHIC with an integrated luminosity of 72~pb$^{-1}$ at $\sqrt{s}=510$~GeV and an average beam polarization of 56\%.

\end{abstract}

\section{Introduction}

Understanding the spin structure of the proton is a fundamental and long standing challenge in nuclear physics.  The contribution of the quark spins to that of the proton have been measured in polarized deep-inelastic scattering (pDIS) experiments, and were found to contribute only $\sim$30\% to the total proton spin \cite{Filippone:2001}.  While the total quark polarization is well determined from inclusive pDIS, flavor separation is accessible only through semi-inclusive measurements which rely on the use of fragmentation functions to relate measurements of the final-state hadrons to the quark and antiquark distributions.  The extracted antiquark helicity distributions have considerably larger uncertainties compared to the well-constrained quark + antiquark sums~\cite{DSSV}.  

The decomposition of quark and antiquark spin contributions by flavor is of interest in its own right, but it has garnered increased attention due to measurements in the unpolarized sector, most recently by the E866/NuSea Drell-Yan experiment.  The E866/NuSea experiment has shown an excess of $\bar{d}$ over $\bar{u}$ quarks in the proton, which has become known as the ``flavor asymmetry'' of the antiquark sea (see Ref.~\cite{Garvey:2001yq} for a review).  This was not anticipated in previous QCD predictions, and brought into question some of the assumptions about the perturbative origin of these sea quarks in the proton.  Several models have been proposed which are consistent with the E866/NuSea data.  Measurements of the flavor separated antiquark spin contributions at RHIC will provide additional constraints to test these models, and give new insight into the mechanism for generating the light quark sea. 

The $u$ and $d$ quark and antiquark distributions are probed at RHIC through the production of $W^{+(-)}$ bosons, for which the dominant process is the fusion of $u+\bar{d}(d+\bar{u})$ quarks.  The parity-violating nature of the weak production process gives rise to large longitudinal single-spin asymmetries, which provide access to the quark and antiquark helicity distributions.  The asymmetry is defined as $A_L = \Delta\sigma/\sigma$, where $\Delta\sigma = \sigma_+ - \sigma_-$ is the difference in the cross section between positive and negative helicity polarized proton beam, and $\sigma$ is the total cross section.
First measurements of the cross section and spin asymmetry, $A_L$, for $W$ production were reported by the STAR~\cite{PhysRevLett.106.062002,PRDstar} and PHENIX~\cite{PhysRevLett.106.062001} collaborations from data collected in 2009 with an integrated luminosity of $\sim 12$ pb$^{-1}$ at $\sqrt{s}=500$~GeV and an average beam polarization of 39\%.  STAR also reported the first measurement of the unpolarized cross section ratio $\sigma_{W+}/\sigma_{W-}$~\cite{PRDstar}, which is sensitive to the proton's unpolarized antiquark distributions and are complementary to measurements of the $\bar{d}/\bar{u}$ ratio in Drell-Yan~\cite{Garvey:2001yq}.   The results presented in this contribution are from data collected in 2012 by the STAR experiment with an integrated luminosity of 72~pb$^{-1}$ at $\sqrt{s}=510$~GeV and an average beam polarization of 56\%.

%Run 12 results and discussion 
\section{Analysis and Results}

% General and mid-rapidity selection
$W \rightarrow e\nu$ candidate events are selected based on differences in the event topology between leptonic $W$ decays and QCD hard-scattering events, such as di-jets, or $Z \rightarrow e^+e^-$ events.  Near mid-rapidity $W \rightarrow e\nu$ events are characterized by an isolated $e^{\pm}$ with a transverse energy, $E_T^e$, that peaks near half the $W$ mass ($\sim$40 GeV) with a distribution referred to as a Jacobian peak.  Leptonic $W$ decays also contain a neutrino, close to opposite in azimuth of the decay $e^{\pm}$, which is undetected and leads to a large missing transverse energy in the event.  As a result, there is a large imbalance in the vector $p_T$ sum of all reconstructed final state objects for $W$ events.  In contrast, $Z \rightarrow e^+e^-$ and QCD background events are characterized by a small magnitude of this vector $p_T$ sum.  Both isolation and "$p_T$-balance" cuts are imposed to suppress those background processes and enhance the $W$ signal contribution (for a more detailed discussion see Ref.~\cite{PRDstar}). 

\begin{figure}
	\begin{center}
    		\includegraphics[width=0.7\textwidth]{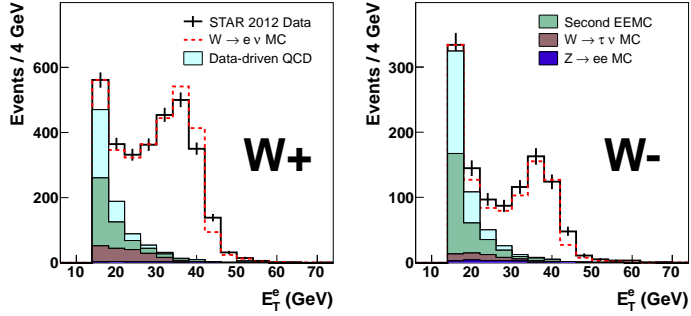} %0.95
    	\caption{  $E_T^e$ distribution of $W^+$ (left) and $W^-$ (right) candidate events (black), background contributions, and sum of backgrounds and $W \rightarrow e\nu$ MC signal (red-dashed). }
    	\label{Fig:figure1}
  	\end{center}	
\end{figure}

%Background estimation and uncertainty
The charge-separated $W^{\pm}$ candidate yields for $e^{\pm}$ pseudorapidity $|\eta_e|<1$ are shown in figure~\ref{Fig:figure1} as a function of $E_T^e$.  Background contributions to the candidate yields were estimated and are shown in figure~\ref{Fig:figure1}, as well.  $W \rightarrow \tau\nu$ and $Z \rightarrow e^+ e^-$ electroweak backgrounds were estimated with PYTHIA Monte-Carlo (MC) samples, while background contributions associated with QCD processes were estimated through two data-driven procedures discussed in more detail in Ref.~\cite{PRDstar}.   The measured $E_T^e$ distribution is in good agreement with the summed background and $W \rightarrow e\nu$ MC (red-dashed line) yields.

% Forward rapidity selection
In this contribution we also present results for $W$s produced with $|\eta_e|>1$.  In this region the Endcap Electromagnetic Calorimeter (EEMC)~\cite{EEMC} is used to reconstruct the $e^{\pm}$ candidate energy, while the Time Projection Chamber provides the charged particle reconstruction similar to the mid-rapidity analysis described in Refs.~\cite{PhysRevLett.106.062002,PRDstar}.  Similar isolation and $p_T$-balance requirements were used to select $W \rightarrow e\nu$ signal events as the mid-rapidity analysis.  Additional background suppression is provided by the EEMC Shower Maximum Detector which allows for a measure of the transverse profile of the $e^\pm$ electromagnetic shower in the EEMC.  For $W \rightarrow e \nu$ decays the $e^\pm$ shower should be isolated with a narrow transverse profile, while QCD background candidates often contain extra energy deposits away from the $e^\pm$ track leading to a wider shower  profile.

\begin{figure}
	\begin{center}
    		\includegraphics[width=0.40\textwidth]{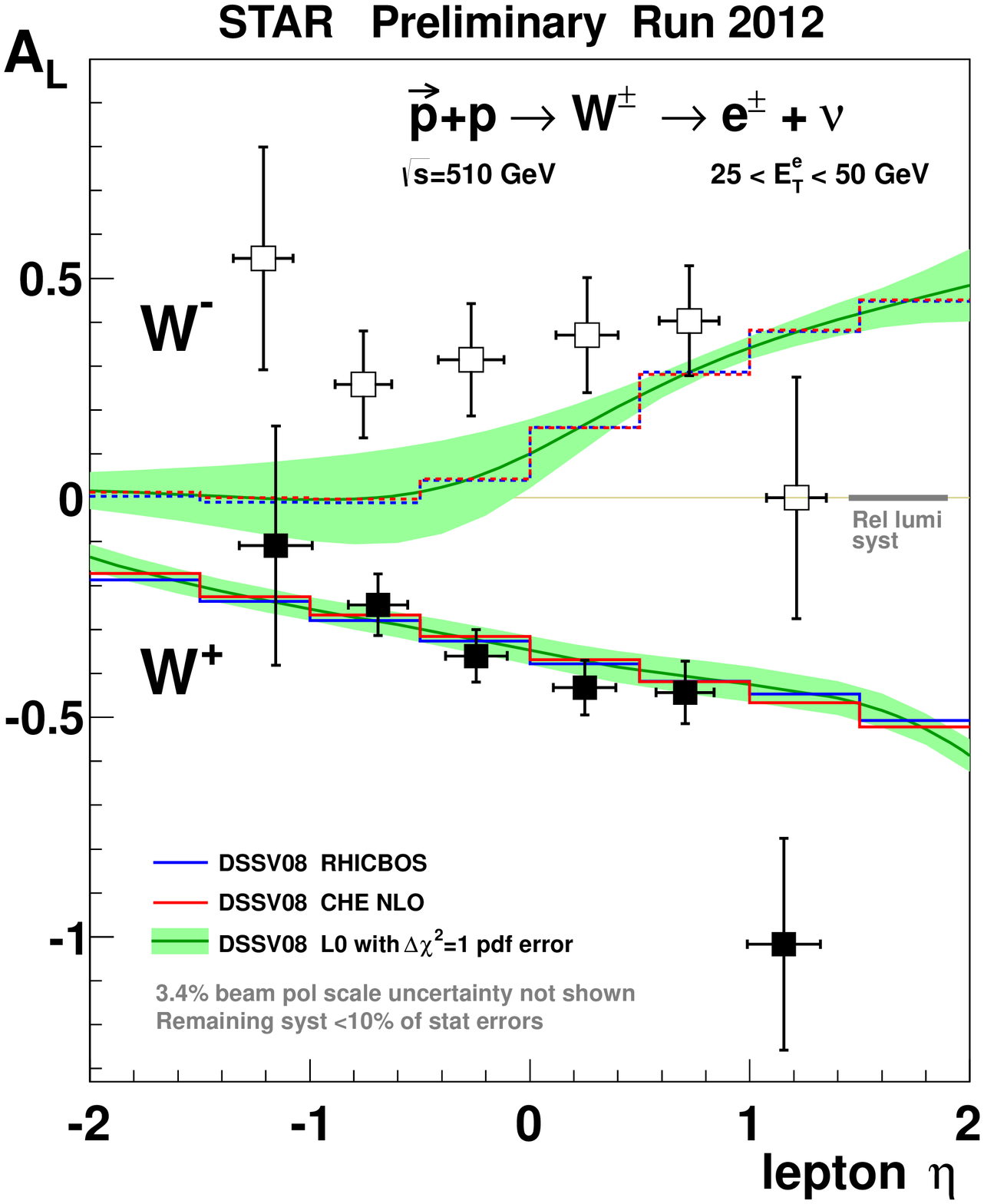} %0.5
    	 	\includegraphics[width=0.33\textwidth]{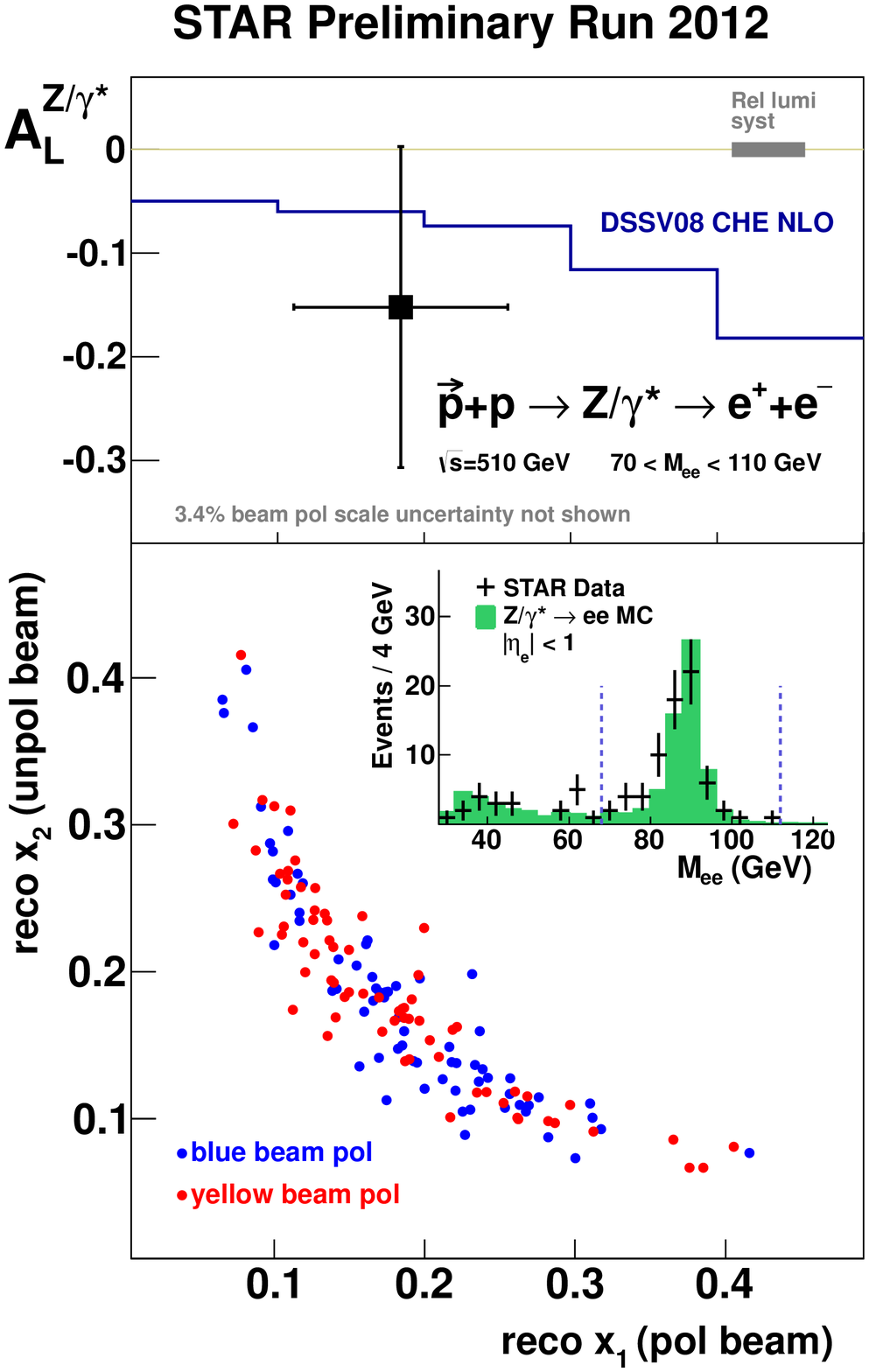} %0.415
    	\caption{ (a) Single-spin asymmetry $A_L$ for $W^\pm$ production as a function of $\eta_e$.  (b) Single-spin asymmetry, $A_L$ for $Z/\gamma^*$ production as a function of reconstructed $x_1$ and correlation of reconstructed $x_1$ and $x_2$ with the $Z/\gamma^*$ invariant mass distribution inset.}
    	\label{Fig:figure2}
  	\end{center}	
\end{figure}

%W AL results
Using a single polarized proton beam, and selecting the charge of the measured $e^\pm$, the parity-violating asymmetry $A_L^W$ can be written as:

\begin{equation}
A_L^{W}(\eta_e) = \frac{1}{P}\frac{1}{\beta}\frac{R_{+}N_{+}^{W}(\eta_e) - R_{-}N_{-}^{W}(\eta_e)}{R_{+}N_{+}^{W}(\eta_e) + R_{-}N_{-}^{W}(\eta_e)} + \frac{\alpha(\eta_e)}{\beta(\eta_e)}
\end{equation}

\noindent for a given $e^\pm$ pseudorapidity, $\eta_e$, where $P$ is the beam polarization, $N_{+(-)}^{W}$ is the $W$ candidate yield for positive (negative) proton beam helicity, and $R_{+(-)}$ are the respective relative luminosities.  The dilution of $A_L$ due to unpolarized background contributions to the $W$ candidate yield are represented by $\beta$.  Polarized background contributions are corrected by the $\alpha$ term which is estimated by the fraction of $Z \rightarrow e^+e^-$ contamination and the expected asymmetry for $Z$ production~\cite{CHE}.  Both proton beams at RHIC are polarized however, so the average asymmetry for the two beams is measured and the results are shown in figure~\ref{Fig:figure2}, with the error bars indicating only statistical uncertainties.  The measured asymmetries have a common 3.4\% normalization uncertainty due to the uncertainty in the measured beam polarization, and the remaining systematic uncertainties due to background and relative luminosity are less than 10\% of the statistical errors. 

The measured asymmetries are compared to theoretical predictions from NLO (CHE)~\cite{CHE} and fully resummed (RHICBOS)~\cite{RHICBOS} calculations based on the DSSV08~\cite{DSSV} helicity distributions.  The predictions for $A_L^{W^+}$ are in good agreement with measured values.  For $A_L^{W^-}$ however, the measured asymmetries are systematically larger than the theoretical predictions.  The electron pseudorapidity region $\eta_{e^-} < -0.5$ in particular is sensitive to the $\bar{u}$ quark helicity distribution, and the enhancement of $A_L^{W^-}$ there indicates a preference for a sizable positive $\bar{u}$ quark polarization relative to that expected from the global analysis of pDIS data.  Indeed, preliminary results from the DSSV++ global analysis of pDIS and RHIC data (reported in Ref.~\cite{spinWriteup}) show a reduced uncertainty in the antiquark polarization and a preference for positive $\Delta\bar{u}$.

%Z results
From the same dataset $Z \rightarrow e^+e^-$ events were also identified by selecting a pair of isolated, oppositely charged $e^\pm$ candidates, as described in Ref.~\cite{PRDstar}.  The resulting invariant mass distribution of $e^+ e^-$ pairs is shown in an inset of figure~\ref{Fig:figure2} with a comparison to the MC expectation.  While $Z$ production is limited at RHIC energies by its small production cross section, one unique advantage is the fully reconstructed $e^+ e^-$ final state, as opposed to the undetected neutrino in the $W$ case.  This allows the initial state kinematics to be measured event by event at leading order via the relation, $x_{1(2)}~=~\left(M_{ee}/\sqrt{s}\right)e^{\pm y_Z}$.  The correlation of the reconstructed $x_1$ and $x_2$ is shown in figure~\ref{Fig:figure2} as well as the single-spin asymmetry for $Z$ production, $A_L^{Z}$, compared to the theoretical prediction as a function of $x_1$, the Bjorken-$x$ of the polarized beam.

\section{Conclusion}

The study of $W$ production in polarized $p+p$ collisions offers a clean and unique probe to further constrain our understanding of the polarized sea quark distributions.  This contribution presents the first measurement of the $\eta_e$ dependence of the single-spin asymmetry $A_L$ for $W^\pm$ production.  These results have been included in a preliminary global analysis of the DSSV group reported in Ref.~\cite{spinWriteup}, which demonstrate the new constraints on the antiquark helicity distributions.  RHIC is planning another longitudinally polarized $p+p$ run at $\sqrt{s}=510$~GeV in 2013 where STAR will collect increased statistics to further constrain the flavor separated antiquark contributions to the proton spin.

\end{document}